\documentclass[12pt]{article}

\newcommand{\ket}[1]{|#1\rangle}
\newcommand{\bra}[1]{\langle #1 |}

\title{Quantum Stabilizer Codes Can Realize
  Access Structures Impossible by Classical
  Secret Sharing}

\author{Ryutaroh Matsumoto\\
  Dept.\ of Information and Communications Engineering\\
  Tokyo Institute of Technology, Japan.\\
  Email: ryutaroh@rmatsumoto.org}
\date{January 11, 2017}

\begin{document}
\maketitle
\begin{abstract}
We show a simple example of a secret sharing scheme
encoding classical secret to quantum shares
that can realize an access structure impossible
by classical information processing with limitation on
the size of each share.
The example is based on quantum stabilizer codes.
\end{abstract}

\section{Introduction}
Secret sharing (SS) \cite{shamir79} is a cryptographic scheme to
encode a secret to multiple shares being distributed to
participants, so that only qualified sets of participants
can reconstruct the original secret from their shares.
Traditionally both secret and shares were classical information
(bits). Several authors \cite{cleve99,gottesman00,smith00}
extended the traditional SS to quantum one
so that a quantum secret can be encoded to quantum shares.
The family of qualified sets are called the access structure.
In this note, we study perfect secret sharing, in which
every unqualified set has absolutely no information about
secret.
An $(n,k)$ threshold secret sharing scheme
distributes secret to $n$ participants and
enables $k$ or more participants to reconstruct
the secret while $k-1$ or less participants have
no information about the secret.

Cascudo et al.\ \cite{cascudo13} studied what kind
of access structures can be realized with small share
size $\bar{q}$, and proved that $\bar{q}\geq n-k+2$ as a special
case of their main result for secret sharing schemes
with classical secret and classical shares,
where $\bar{q}$ is the average of the cardinalities of
all share sets. This means that there cannot be a $(3,5)$
threshold scheme distributing 1-bit secret to
5 participants receiving 1-bit shares,
because $\bar{q}=2$, $n=5$ and $k=3$ in that case.

For secret sharing schemes with quantum secret and
quantum shares, it has been observed that quantum
error-correcting codes (QECC) can be used for the purpose
\cite{cleve99,gottesman00,marin13,matsumoto14qss,smith00}.
A quantum secret is simply encoded by a QECC
into $n$ quantum systems, which are distributed
to $n$ participants.
A qualified set of participants can reconstruct the quantum secret
by the erasure-correcting procedure  with treating missing shares
as erasures.
The dimension of a quantum share corresponds to the size of
a classical share, for example, a qubit corresponds to
a classical share of size 2, i.e., a bit.

The above QECC can also
distribute classical secret, just by regarding classical
secret (random variable) as quantum information (density matrix).
The purpose of this note is to show that the well-known
binary stabilizer QECC can realize a $(3,5)$
secret sharing scheme distributing 1-bit classical
secret to 5 participants receiving 1-\emph{qubit}
of quantum shares.
If we regard a qubit as an equivalent information unit to a bit,
it means $q=2$, $n=5$ and $k=3$ that cannot be realized by
purely classical information processing \cite{cascudo13}.
The rest of this note mainly consists of an analysis of
the access structure of a secret sharing scheme
constructed from a binary stabilizer QECC.

We note that a $(3,5)$ scheme was already realized by
\cite[Section IV.B]{markham08}
with qubit shares, but the QECC based secret sharing
has a much simpler protocol because it just uses an
encoding procedure and the erasure decoding procedure of the QECC
(or the recently proposed unitary reconstruction procedure \cite{matsumoto17uni}).

\section{Access Structure of the Secret Sharing Scheme}
\subsection{Secret Sharing Scheme and Qualified Sets}
We consider a binary stabilizer QECC encoding
1 qubit to 5 qubit \cite[Section 3.3]{gottesmanthesis}.
Qubit $\ket{0}$ is encoded to
\begin{eqnarray*}
&& \ket{\psi(0)} \\
&= & \ket{00000} + \ket{10010} + \ket{01001} + \ket{10100}  \\
& & \mbox{} + \ket{01010} - \ket{11011} - \ket{00110} - \ket{11000} \\
& & \mbox{} - \ket{11101} - \ket{00011} - \ket{11110} - \ket{01111} \\
& & \mbox{} - \ket{10001} - \ket{01100} - \ket{10111} + \ket{00101}, 
\end{eqnarray*}
and $\ket{1}$ is encoded to
\begin{eqnarray*}
&& \ket{\psi(1)} \\
& = & \ket{11111} + \ket{01101} + \ket{10110} + \ket{01011}  \\
& & \mbox{} + \ket{10101} - \ket{00100} - \ket{11001} - \ket{00111}  \\
& & \mbox{} - \ket{00010} - \ket{11100} - \ket{00001} - \ket{10000} \\
& & \mbox{} - \ket{01110} - \ket{10011} - \ket{01000} + \ket{11010}.
\end{eqnarray*}

To share classical bit $0$ (resp.\ $1$),
each qubit in $\ket{\psi(0)}$ (resp.\ $\ket{\psi(1)}$ is distributed to each participants.
In error correction,
an erasure means a (quantum or classical) error with known location.
The above QECC can correct up to two erasures.
By regarding classical bit $0$ as quantum information $\ket{0}\bra{0}$
and $1$ as $\ket{1}\bra{1}$, we can see that three or more
participants can reconstruct the classical secret.

\subsection{Unqualified Sets}
In order to clarify the access structure of
the above secret sharing scheme,
we have to study what can be known to two or less participants.
One can use the Holevo information \cite{chuangnielsen}
between classical information (probability distribution)
and quantum information (density matrix)
to study such a problem.
Let $J \subset \{1$, \ldots, $5\}$ be a set of
shares (participants),
and $\rho_s^J$ be the quantum state of
shares in $J$ corresponding to secret $s \in \{0$, $1\}$.
Let $q_0$ and $q_1$ be a probability distribution of
the secret $s$ on $\{0,1\}$.
The Holevo information is defined by
\[
I(J) = S(q_0 \rho_0^J + q_1 \rho_1^J) -
( q_0 S( \rho_0^J) + q_1 S( \rho_1^J)).
\]
If $I(J)=0$ then  outcomes (interpreted as random variables)
of any measurement of
the share set $J$ are statistically independent of
the secret $s$.

The secret sharing scheme considered here
can also be regarded as sharing quantum
secret $\alpha_0 \ket{0} + \alpha_1 \ket{1}$
producing the quantum shares $\alpha_0 \ket{\psi(0)} + \alpha_1\ket{\psi(1)}$.
If $|J| \leq 2$ then
$\overline{J} = \{1$, \ldots, $5\} \setminus J$
can reconstruct the quantum secret $\alpha_0 \ket{0} + \alpha_1 \ket{1}$,
as the minimum distance of this QECC is 3 \cite{gottesmanthesis}.
By \cite{cleve99,gottesman00}, this implies that
$J$ has absolutely no information about the quantum
secret $\alpha_0 \ket{0} + \alpha_1 \ket{1}$.
By \cite[Theorem 2]{ogawa05}, we see $I(J)=0$,
which means that $J$ has no information about the classical secret $s$,
in other words,
any measurement on $J$ gives outcomes statistically independent of $s$.

So we can see that this QECC is a $(3,5)$ scheme distributing one qubit
to each participant as a share.
If each qubit in a share is replaced by a bit,
then a $(3,5)$ scheme cannot be realized \cite{cascudo13}.
Thus this simple example demonstrates that
an impossible access structure given a limitation on the size of shares
within classical information processing
sometimes becomes realizable by quantum information processing,
especially by quantum stabilizer codes.
This is a stark contrast to the fact that quantum secret sharing
constructed from the CSS QECC \cite{calderbank96,steane96}
has an access structure whose qualified sets are always
qualified \cite{matsumoto14qss} in the corresponding classical secret sharing.

\section*{Acknowledgment}
The author would like to thank Prof.\ Tomohiro Ogawa for
helpful discussion.
This research is partly supported by the JSPS Grant  
 No.\ 26289116.


\end{document}